\documentclass[useAMS,usenatbib]{mn2e}
\usepackage{epsfig}
\usepackage{IEEEtrantools}
\usepackage{multirow}
\usepackage{natbib}
\usepackage{hyperref}
\usepackage{caption}
\usepackage{subcaption}
\usepackage{comment}
\usepackage{mathrsfs}
\usepackage{amsmath}

\usepackage{amssymb}
\hypersetup{colorlinks=true, citecolor=blue, linkcolor=red,pdfborder=0 0 1,citebordercolor={0 1 0}}
\usepackage[toc,page]{appendix}
\usepackage{graphicx}

\usepackage{xspace}
\def \inte {{\em INTEGRAL}}
\def \intejx {{\em INTEGRAL/JEM-X}}
\def \inteis {{\em INTEGRAL/ISGRI}}
\def \swift {{\em Swift}}
\def \beppo{{\em BeppoSAX}}
\def \fermigbm{{\em FERMI-GBM}}
\def \chandra {{\em Chandra}}
\def \xmm {{\em XMM-Newton}}

\def \gro{{\em GRO J1744$-$28}}

\def \swiftxrt{{\em Swift-XRT}}
\def \swiftbat{{\em Swift-BAT}}

\title[Accretion torque in GRO J1744--28]{Study of the accretion torque during the 2014 outburst of the X-ray pulsar GRO J1744--28 }

\author[Sanna et al. ]{A. Sanna$^{1}$\thanks{E-mail:
    andrea.sanna@dsf.unica.it}, A. Riggio$^{1}$, L. Burderi$^{1}$, F. Pintore$^{2}$, T. Di Salvo$^{3}$, A. D'A\`i$^{4}$,  \newauthor
    E. Bozzo$^{5}$, P. Esposito$^{6}$, A. Segreto$^{7}$, F. Scarano$^{1}$, R. Iaria$^{3}$, A.~F. Gambino$^{3}$\\
$^{1}$Dipartimento di Fisica, Universit\`a degli Studi di Cagliari, SP Monserrato-Sestu km 0.7, 09042 Monserrato, Italy\\
$^{2}$INAF-Istituto di Astrofisica Spaziale e Fisica Cosmica - Milano, via E. Bassini 15, I-20133 Milano, Italy\\
$^{3}$Universit\`a degli Studi di Palermo, Dipartimento di Fisica e Chimica, via Archirafi 36, 90123 Palermo, Italy\\
$^{4}$INAF/IASF Palermo, via Ugo La Malfa 153, I-90146, Palermo, Italy\\
$^{5}$ISDC, Department of astronomy, University of Geneva, chemin d'\'Ecogia, 16 CH-1290 Versoix, Switzerland\\
$^{6}$Anton Pannekoek Institute for Astronomy, University of Amsterdam, Postbus 94249, NL-1090 GE Amsterdam, The Netherlands\\
$^{7}$Istituto di Astrofisica Spaziale e Fisica Cosmica - Palermo, INAF, via U. La Malfa 153, 90146 Palermo, Italy}
\begin{document}

\date{Accepted 2017 March 10. Received 2017 March 9; in original form 2016 October 17}

\pagerange{\pageref{firstpage}--\pageref{lastpage}} \pubyear{}

\maketitle

\label{firstpage}

\begin{abstract}
We present the spectral and timing analysis of the X-ray pulsar \gro{} during its 2014 outburst using data collected with the X-ray satellites \swift{}, \inte{}, \chandra{}, and \xmm{}. We derived, by phase-connected timing analysis of the observed pulses, an updated set of the source ephemeris. We were also able to investigate the spin-up of the X-ray pulsar as a consequence of the accretion torque during the outburst. Relating the spin-up rate and the mass accretion rate as $\dot{\nu}\propto\dot{M}^{\beta}$, we fitted the pulse phase delays obtaining a value of $\beta=0.96(3)$. Combining the results from the source spin-up frequency derivative and the flux estimation, we constrained the source distance to be between 3.4--4.1 kpc, assuming a disc viscous parameter $\alpha$ to be in the range 0.1--1. Finally, we investigated the presence of a possible spin-down torque by adding a quadratic component to the pulse phase delay model. The marginal statistical improvement of the updated model does not allow us to firmly confirm the presence of this component.  
\end{abstract}

\begin{keywords}
accretion, accretion disc, --stars: neutron --X-rays: binaries --X-rays: individuals: GRO J1744--28
\end{keywords}

\section{Introduction}
Accretion powered X-ray pulsars in low mass X-ray binaries (LMXB) are magnetised neutron stars (NS) that accrete via Roche-lobe overflow. Because of the high specific angular momentum carried by the transferred matter, an accretion disc is formed. Viscous stresses allow the accretion disc to lose angular momentum and to extend in the vicinity of the NS where the disc plasma starts interacting with the NS magnetic field. The matter captured by the pulsar magnetosphere is forced to follow the magnetic field lines and accretes onto the  polar caps of the NS. Therefore, a net torque is exerted on the compact object causing a variation of its spin frequency. This is the fundamental process at the base of the so called \textit{recycling scenario} \citep[see, e.g.,][]{Bhattacharya91} that describes the millisecond radio pulsars as the final stage of a long-lasting process of accretion of matter onto an initially slowly-spinning NS hosted in a LMXB.
Accreting X-ray pulsars offer the unique possibility to directly investigate the interaction between the accretion disc matter and the NS magnetic field by means of the study of their spin evolution.  

The X-ray source \gro{} was firstly discovered in the hard energy domain (up to $\sim$75 keV) by the \textit{Burst and Transient Source Experiment} (BATSE) aboard the {\em Compton Gamma Ray Observatory} satellite on 1995 December 2 \citep{Kouveliotou96} as a bursting source, showing very rapidly recurrent X-ray bursts explained later on as spasmodic accretion episodes \citep[e.g., ][]{Kouveliotou96, Lewin96}.  Soon after the discovery, a coherent pulsation at 467 ms was detected \citep{Finger96}, making \gro{} the first (at the time) X-ray pulsating burster (for this reason renamed as the \textit{Bursting Pulsar}).
 \citet{Finger96} placed the source in a binary system with an orbital period of about 11.83 days and a mass function of $1.36 \times 10^{-4}$ M$_{\odot}$. The small value of the mass function indicates that the secondary is likely a low-mass star \citep[$\sim 0.2$ M$_\odot$, see e.g., ][]{Daumerie96, Lamb96, vanparadijs97} or that the inclination of the orbit is relatively small ($< 20^\circ $).
The distance of the source is currently unknown, however the heavy absorption towards the source (equivalent hydrogen column density $N_{H}\approx10^{23}$ cm$^{-2}$) seems to place \gro{} towards the Galactic centre \citep[][]{Fishman95,Kouveliotou96}, at a distance of about 8 kpc. Studies on the NIR counterparts of the source, however, point towards a smaller value of about 4 kpc \citep[][]{Gosling07,Wang07}

On January 2014, after almost two decades of quiescence state, \gro{} re-activated in X-rays \citep[see e.g.,][]{Negoro14, Kennea14}. Soon after, from the same direction of the sky, \citet{Finger14} detected a 2.14 Hz pulsation with \fermigbm{} monitor, confirming the beginning of a new outburst. The presence of X-ray pulsations has been confirmed with \swiftxrt{} \citep[][]{Dai14}, which regularly monitored the evolution of the outburst. After almost 100 days from the beginning of the outburst ($\sim$ 56790 MJD) the source fainted out entering in a quiescence state.

Before the 2014 outburst, the magnetic field of the source resulted undetermined, however, several constrains have been set up. \citet{Finger96}, based on the spin-up rate of the source, set an upper limit on the dipole magnetic field of $B \leq 6 \times10^{11}$ G. \citet{Cui97} estimated a surface magnetic field of $B\approx2.4\times 10^{11}$ G. This value was deduced by invoking the propeller effect to explain the lack of pulsation below a certain flux threshold. \citet{Rappaport97}, from binary evolution calculations, constrained the dipole magnetic field of the source in the range of $(1.8-7)\times 10^{11}$ G. \citealt{DAi15}, during the latest outburst of the source, detected a cyclotron resonant scattering feature centered at $\sim 4.7$ keV, from which they inferred a dipole magnetic field of $B \simeq 5.27\times 10^{11}$ G. These result has been later on confirmed by the analysis of \beppo{} observation of \gro{} from the 1997 outburst \citep{Doroshenko15}.

Here we present the results of the timing analysis of the 2014 outburst of \gro{}. On Section 2 we describe the observations and the relative data reduction. Section 3 describes the timing techniques. Finally on Section 4 we discuss the results of the analysis and their implication in the context of accretion onto magnetised NS. 
 
\section{Observations}

In this work we consider the whole set of observations of \gro{} collected with \swift{}, \inte{}, \chandra{}, and \xmm{} between 2014 February 9 (almost 20 days after the first detection reported by \fermigbm) and 2014 May 25 (MJD 56696.0--56803.0), when the source was detected from these observatories, as shown in Fig.~\ref{fig:flux}. Several X-ray bursts have been detected during the outburst. Here we focus on the timing properties of the source during the persistent emission (non-bursting phases), therefore we discard each burst significantly detected above the continuum emission, excluding data within a time interval between 25 and 60 seconds centred on the burst peak, depending on the burst duration.

\subsection{Swift}
\subsubsection{XRT}
The \swift{} observatory monitored \gro{} with the X-ray Telescope \citep[XRT,][]{Burrows2005a} starting from MJD 56702.58 up to MJD 56846, for a total of $\sim 100$ ks of data. However, the source was significantly detected by \swiftxrt{} only up to MJD 56803. We extracted and reprocessed the data using the standard procedures and the latest calibration files (June 2014; CALDB v. 20140605). Data have been processed with the \textsc{xrtpipeline v.0.12.9}. Timing analysis has been carried out using data collected in window timing (WT) mode in the energy range 0.3--10 keV and characterised by time resolution of 1.7 ms. To perform the spectral analysis we used both the WT and photon counting (PC) mode data in the 0.3--10 keV range. The source counts were extracted within a 30-pixel radius (corresponding to $\sim71''$), while the background counts were extracted within an annular region with inner and outer radius of 33 and 46 pixels, respectively. Given the relatively high count rate observed in the WT mode observations at the peak of the outburst ($\sim 80$ cts s$^{-1}$), we decided to investigate possible pile-up contaminations. For these WT observations with the highest count rates we extracted a source spectrum from an annular region for which, starting from zero value, we increased progressively the inner radius in pixels. Fitting the spectra with the same model did not show any significant discrepancy between the model parameters, suggesting no pile-up contamination. Moreover, for the same extraction region described before, we checked the distribution of different event grades (from 0 to 2), finding differences in the distribution smaller than 1$\%$. We concluded that the \swiftxrt{} observations were not affected by pile-up. Event files from WT mode observations have been barycentred with the tool \texttt{barycorr}, using the coordinate of the source (see Section~\ref{sec:timing}). 

\subsubsection{BAT}
The \swiftbat{} survey data, retrieved from the HEASARC public
archive (\url{http://swift.gsfc.nasa.gov/archive/}), were processed using 
\texttt{BAT-IMAGER} software \citep{Segreto10}.
This code, dedicated to the processing of coded mask instrument data,
computes all-sky maps  and, for each detected source, produces 
light curves and spectra.

\subsection{\inte{}}
\inte{} data were analysed using version 10.0 of the OSA
software distributed by the ISDC \citep{courvoisier03}.
\inte{} observations are divided in ``science windows'' (SCWs),
i.e. pointings with typical durations of $\sim$2~ks.
We considered all the publicly available SCWs for the IBIS/ISGRI
\citep[17-80~keV,][]{lebrun03,ubertini03} that were performed during the
outburst of the source, including observations of the
Galactic bulge carried out in satellite revolution
1384, 1386, 1389, 1392, 1395, 1398, 1401, 1404, and 1407.
We also used of the Galactic center region observed in revolution
1386.
In all these observations, the source was within 12$^\circ$ from the center of the ISGRI
field of view (FoV), thus avoiding problems with any instrument calibration uncertainties.
We also extracted JEM-X1 and JEM-X2 \citep[][]{lund03} data every time the source was within
the FoV of the instrument. We note that no JEM-X data were available during revolution 1389 due to
a solar flare that forced the instrument in safe mode.
A light-curve of the source was extracted from the JEM-X units at 2\,s
resolution in order to identify type-I X-ray bursts. These were removed during the extraction of the
ISGRI and JEM-X persistent spectra by creating manually all the required good-time-intervals files.
We rebinned the ISGRI (JEM-X) response matrix in order to have 37 (32) bins
spanning the energy range 20-180~keV (3-35~keV) for all spectra. This maximised the signal-to-noise (S/N)
of the data. The source events files were extracted for ISGRI and JEM-X in each SCWs by using the
\textsc{evts\_extract} tool distributed with OSA. The \inte{}, burst-filtered, event files have been barycentred with respect to the Solar system barycentre using the tool \texttt{barycent}, using the coordinate of the source (see Section~\ref{sec:timing}).
\subsection{XMM-Newton}
\xmm{} observed \gro{} on 2014 March 6 for a total of $\sim81$ ks of data. The EPIC-pn \citep[hereafter PN;][]{Strueder01} operated in Timing Mode with the optical thick filter, while the Reflection Grating Spectrometer instruments (RGS1 and RGS2) in Spectroscopy Mode. The EPIC-MOS CCD were kept off to allocate the highest possible telemetry for the PN. The PN event file was processed using the \textsc{epproc} pipeline processing task \textsc{rdpha}, as suggested by the most recent calibration \citep[see e.g., Guainazzi et al. 2013\footnote{http://xmm2.esac.esa.int/docs/documents/CAL-SRN-0248-1-0.ps.gz};][]{Pintore14}. Events have been filtered selecting {\sc pattern $\leq$ 4} (allowing for singles and doubles pixel events only), and {\sc `flag=0'}. Source events were extracted within the RAWX range [31:43]. The average count rate, corrected for telemetry gaps (\textsc{epiclccorr} tool) during the entire observation over all the PN CCD, was 714 cps. We barycentred the PN, burst-filtered, data with respect to the Solar system barycentre using the tool \texttt{barycen}, using the coordinate of the source (see Section~\ref{sec:timing}).

\subsection{Chandra}
\chandra{} observed \gro{} three times during the latest outburst. The observations were performed on March, the 3rd 2014 (ObsId. 16596), March the 29th (ObsId. 16605), and March the 31st (ObsId. 16606), for a total exposure time of $\sim 80$ ks.   
\chandra{} data were processed using CIAO 4.7 and CALDB 4.6.5. We obtained clean event level $=$ 2 files using the \texttt{chandra$\_$repro} script. We barycentred the events using the \texttt{axbary} tool, adopting the source coordinates (see Section~\ref{sec:timing}). We extracted from the Type2 PHA event file the High Energy Grating (HEG) spectra according to standard pipeline. 
Spectral response files were created using \textsc{mkgrmf}, and \textsc{fullgarf}.

\begin{figure*}
\centering
\includegraphics[width=1\textwidth]{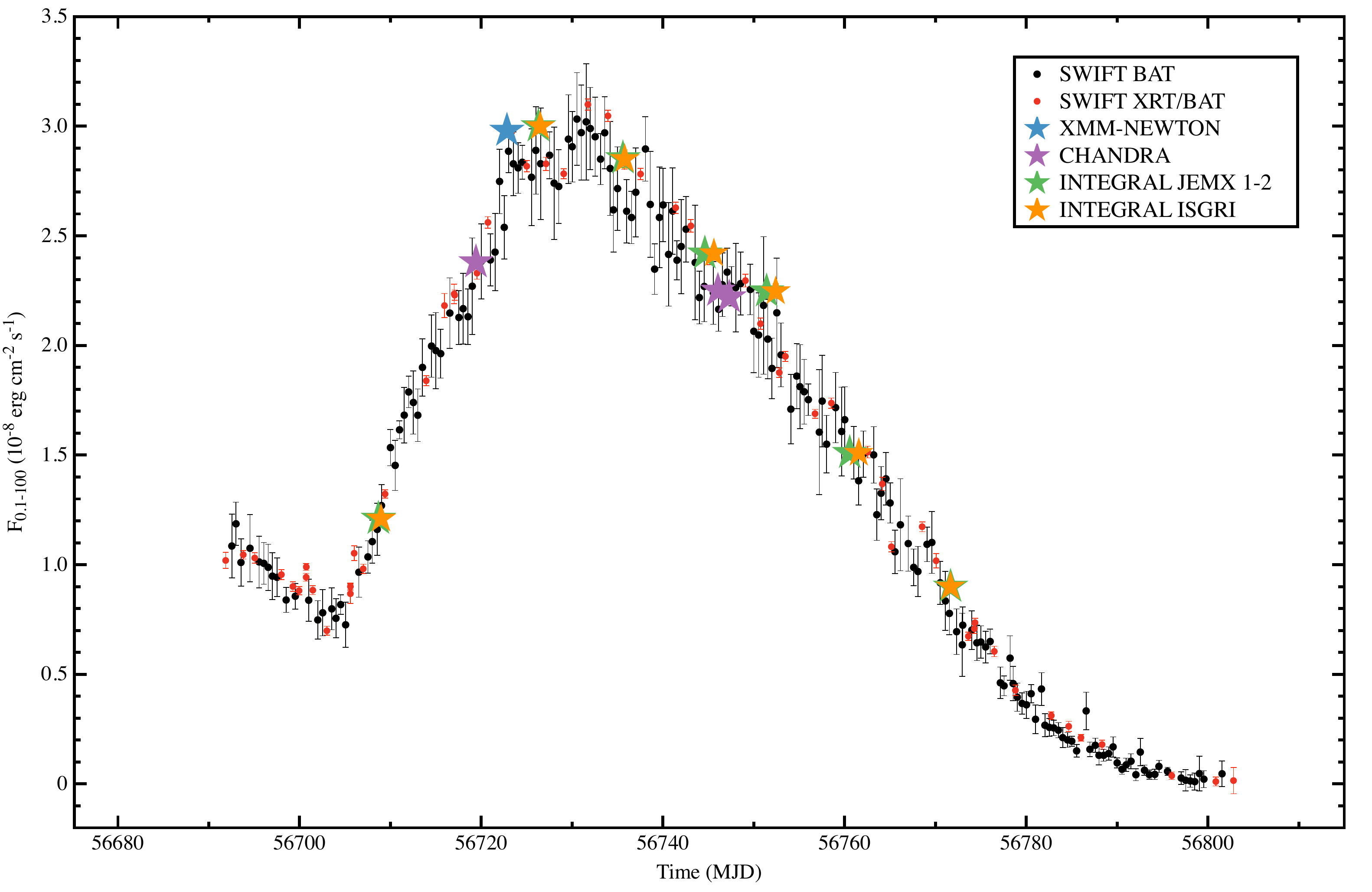}
\caption{Unabsorbed flux measurements of \gro{} estimated by extrapolating the spectral model, \textit{phabs(cutoffpl+gaussian)}, to the 0.1--100 keV energy range (see text). Black points represent the observations performed by \swiftbat{}, red points refer to simultaneous \swiftxrt{} and \swiftbat{}. Stars represent observations collected by \xmm{} (blue), \chandra{} (violet), \intejx{} (green) and \inteis{} (orange), respectively. We note that the unabsorbed flux estimates derived from six different instruments onboard on 4 X-ray satellites, show a good agreement with an accuracy level $\leq20\%$}
\label{fig:flux}
\end{figure*}

\section{Data analysis and results}

\subsection{Spectral analysis}

To have an accurate estimate of the bolometric flux and its evolution during the 2014 outburst of \gro{}, we investigated the broad-band energy spectrum collecting the data from the X-ray instruments that observed the source. The \swift{} all-sky monitor combined with the pointing observations allowed us to have an almost daily coverage of the source in the broad-band energy range 0.3--70 keV. Therefore, when available, we fitted the simultaneous \swiftxrt{} and \swiftbat{} observations of the source, in the energy range 0.3--10 keV and 25--70 keV, respectively. The data were well-fitted (with reduced $\chi^2$ ranging between 0.8 and 1.4 for $\sim$ 880 degrees of freedom) with a simple model consisting of an absorbed cut-off power-law (\texttt{constant*phabs*cutoffpl} on Xspec), adopting the abundance table of \citet{Anders89}, and setting the cross-section table to that of \citet{Balucinska-Church92}. The average parameters were: $N_H = 5.4 \times 10^{22}$ atoms cm$^{-2}$, photon index $\Gamma \sim 0.1$, and high energy cut-off $\sim 7.5$ keV. Occasionally, we found the need to add a broad gaussian emission feature (average centroid energy $E \sim 6.7$ keV, and broadness $\sigma \sim 0.6$ keV). Given the low statistic of the \swiftbat{} observations, we held the normalisation factors between \textit{XRT} and \textit{BAT} fixed to 1. 

To model the spectrum of the \swiftbat{} observations where no simultaneous \swiftxrt{} were available, we applied the model described above fixing $N_H$ and $\Gamma$ to the values found in the nearest \swiftxrt{} observation, while we left the high energy cut-off and the normalisation of \texttt{cutoffpl} free to vary. We then extrapolated the unabsorbed flux in the energy range 0.1--100 keV. In Fig.~\ref{fig:flux} we report the evolution during the outburst of the unabsorbed flux estimated from the simultaneous \textit{Swift-XRT/BAT} (red) and \swiftbat{} observations (black).

To cross-check the spectral model used to describe the \swift{} observations, as well as to improve the outburst coverage, we investigated data collected from other X-ray satellites such as \chandra{}, \inte{} and \xmm{}.
For the spectral analysis of the \chandra{} observations, we fitted the HEG spectra in the energy range 2--8 keV with the model \texttt{phabs*(cutoffpl+gaussian)} limiting the high energy cut-off to the values of the closest simultaneous \textit{Swift-XRT/BAT} observations. All three source spectra were well fitted by the model described above, with values of reduced $\chi^2\sim0.95$ (for $\sim710$ d.o.f.). We estimated the unabsorbed flux in the energy range 0.1--100 keV by extrapolating the model described above. To crosscheck the spectral modelling of these data we compared the unabsorbed flux values of the \chandra{} observations (ObsIds 16605 and 16606) reported by \citet{Degenaar14}. In particular we found the values of the unabsorbed flux in the energy range 2--10 keV of $(9.83\pm0.06)\times 10^{-9}$ erg s$^{-1}$ cm$^{-2}$ (ObsId. 16605), and $(9.76\pm0.07)\times 10^{-9}$ erg s$^{-1}$ cm$^{-2}$ (ObsId. 16606), to be consistent with that already reported. Flux measurements of the \chandra{} observations are shown in Fig.~\ref{fig:flux} in purple.

In order to fit the energy spectra accumulated by \inte{} during the 2014 outburst of \gro{}, for each satellite revolution we extracted the 9--18 keV \textit{JEM-X1}, the 9--25 keV \textit{JEM-X2} and the 25--70 keV \textit{ISGRI} spectra.
 We fitted these data together with the almost simultaneous 0.3--10 keV \swiftxrt{} spectrum, in order to have a better constrain at low energies. 
We modelled the data with the already described \texttt{constant*phabs*(cutoffpl+gaussian)} model, letting the normalisation constants free to vary. 
The data were well-fitted with $\tilde{\chi}^2$ ranging between 0.9 and 1.6 for $\sim$ 860 d.o.f. 
We then estimated the unabsorbed flux in the energy range 0.1--100 keV, that we reported in Fig.~\ref{fig:flux} in green (\textit{JEM-X 1/2}), and orange (\textit{ISGRI}).
Finally, in Fig.~\ref{fig:flux} (blue) we reported the unabsorbed flux of the \xmm{} observation derived by \citet{DAi15}, which modelled the continuum as \texttt{constant*phabs*(diskbb+nthcomp)}. 

We note that the unabsorbed flux estimates extrapolated from the energy range 0.1--100 keV and derived from six different instruments onboard on 4 X-ray satellites, show a good agreement with an accuracy level $\leq20\%$ (see Fig.~\ref{fig:flux}). Compatible results have been reported by \citet{Guver2016a} comparing flux measurements obtained with \xmm{}, \chandra{}, and the {\em Rossi X-ray Timing Explorer}.

\subsection{Timing analysis}
\label{sec:timing}
In order to perform the timing analysis of the 2014 outburst of \gro{},  we used all the collected data. For each satellite, we corrected X-ray photons arrival times (ToA, hereafter) for the motion of the spacecraft with respect to the Solar System Barycentre (SSB), by using spacecraft ephemerides, and adopting the source position derived from a \chandra{} observation taken in the quiescence state \citep[RA=266.137875$^\circ$ DEC=-28.740833$^\circ$, 1$\sigma$ confidence radius of $0.8''$;][]{Wijnands02}. These coordinates are consistent within the errors with the recent source position estimated by \citet[][]{Chakrabarty14} using a \chandra{} observation during the latest outburst (RA=266.137750$^\circ$, DEC=-28.740861$^\circ$, 1$\sigma$ confidence radius of $0.6''$).
We then corrected the ToAs for the delays caused by the binary motion applying the orbital parameters provided by the GBM pulsar team\footnote{http://gammaray.nsstc.nasa.gov/gbm/science/pulsars/} (reported in Table~\ref{tab:par2}) through the recursive formula 
\begin{equation}
\label{eq:barygen} 
t + \frac{z(t)}{c} = t_{arr},
\end{equation}
where $t$ is photon emission time, $t_{arr}$ is the photon arrival time to the SSB, $z(t)$ is the projection along the line of sight of the distance between the NS and the barycenter of binary system, and $c$ is the speed of light. For almost circular orbits ($e \ll 1$) we have:
\begin{equation} 
\label{eq:bary}
\frac{z(t)}{c} = x \left[\sin \ell -e\sin\omega(\cos\ell-1)^2+\frac{e}{2}\cos\omega\sin2\ell \right]
\end{equation}
where $x$=$a_{1} \sin{\textit{i}/c}$ is the projected semimajor axis of the NS orbit in light seconds, $\ell=\ell(t)=\Omega(t-T_{NOD})$ is the NS mean longitude, $\Omega = 2\pi/P_{orb}$, $P_{orb}$ is the orbital period, $T_{NOD}$ is the time of ascending node passage and $w$ is the longitude of the periastron measured from the ascending node. 
The correct emission times (up to an overall  constant $D/c$, where $D$ is the distance between the SSB and the barycenter of the binary system) are calculated by solving iteratively the aforementioned Eq.~\ref{eq:barygen}, $t_{n+1} = t_{arr} - z(t_{n})/c$,
with $z(t)/c$ defined as in Eq.~\ref{eq:bary},
with the conditions $D/c = 0$, and $z(t_{n=0}) = 0$.

\noindent
To compute statistically significant pulse profiles, we split the data into $N$ time intervals of approximately 1000 s length that we epoch-folded in 8 phase bins at the spin frequency $\nu_0 = 2.1411117(4)$ Hz, reported by \citet{Finger14} and corresponding to the time interval 2014 January 21--23.
We fitted each folded profile with a sinusoid of unitary period in order to obtain the corresponding sinusoidal amplitude and the fractional part of the Epoch--Folded Phase Residual. We considered only folded profiles for which the ratio between the amplitude of the sinusoid and its 1$\sigma$ uncertainty was larger than 3.   
We tried to fit the folded profiles including a second harmonic component, but the amplitude of the second harmonic was significant only in a small fraction ($<$5\%) of the folded profiles. 

Before starting the analysis of the pulse phase delays it is important to evaluate any source of error in the observed phase variations. 
To take into account the residuals induced by the motion of the Earth for small variations of the source position $\delta_{\lambda}$ and $\delta_{\gamma}$ expressed in ecliptic coordinates $\lambda$ and $\gamma$, we used the expression:
\begin{equation}
R_{pos}(t) = - \nu_0 y [\sin(M_0+\epsilon)\cos \gamma \delta\lambda -  \cos(M_0+\epsilon)\sin \gamma \delta\gamma],
\end{equation}
where $y=r_E/c$ is Earth's semi-major axis in light-seconds, $M_0=2 \pi (T_0-T_{v})/P_{\oplus}-\lambda$, with $T_{v}$ being the vernal point and $P_{\oplus}$ is the Earth orbital period, $\epsilon=2\pi(t-T_0)/P_{\oplus}$ \citep[see, e.g.,][]{Lyne90}. The short length of our dataset ($\sim$ 90 days) compared with the Earth orbital period does not allow us to disentangle the phase delay residuals induced by uncertainties on the source position with respect to those caused by spin frequency uncertainties or spin frequency derivative. The resulting systematic error in the linear term of the pulse phase delays, which corresponds to an error in computing the spin frequency correction $\delta \nu_0$, can be as $\sigma_{\nu_{pos}}\leq \nu_0y\sigma_{v}(1+\sin^2\gamma)^{1/2}2\pi/P_{\oplus}$, where $\sigma_{v}$ is the positional error circle. On the other hand, the error associated to the quadratic term, that corresponds to an error in computing the spin frequency derivative, can be expressed as $\sigma_{\dot{\nu}_{pos}}\leq \nu_0y\sigma_{v}(1+\sin^2\gamma)^{1/2}(2\pi/P_{\oplus})^2$. Considering the positional uncertainty of $0.8 ''$ reported by \citet{Wijnands02}, we estimated for the 2014 outburst, $\sigma_{\nu_{pos}} \leq 8\times 10^{-10}$ Hz and $\sigma_{\dot{\nu}_{pos}} \leq 1.6\times 10^{-16}$ Hz/s, respectively. These systematic uncertainties will be added in quadrature to the statistical errors of $\nu_0$ and $\dot{\nu}$ estimated from the timing analysis (see Section~\ref{sec:acctor}).
For the specific case of \gro{}, we isolated another phenomenon that can cause temporal delays and that should be taken into account before the timing analysis. During the 1996 outburst, glitches in the arrival times of the X-ray pulses have been observed in correspondence with large X-ray bursts \citep{Stark96}, causing average phase lags of the pulse profile of $\sim$0.03 seconds during each X-ray burst event and recovered on timescales of $\sim1000$ seconds. The origin of this delay is still quite unclear and the attempt to interpret this phenomenon is beyond the scope of this paper. However, the existence of a phase lag with such a long recovery timescale cannot be ignored while investigating variations of the spin frequency of the X-ray pulsar by means of the pulse phase evolution. Since we were not able to predict and correct the phenomenon, we therefore decided to quantify the fluctuations a posteriori and to treat them as a systematic uncertainty proceeding as follows: i) we selected observations longer than 5000 seconds (only 15 observations fulfilled this criterion); ii) we fitted the phase delays with a linear function (under the assumption that quadratic terms are not significant in such a short timescales) and calculated the corresponding data root mean square and the associated statistical error; iii) we calculated a weighted average of the sample observations finding a phase fluctuation, $\sigma_{\phi_{gli}}\sim 1.7\times 10^{-2}$. The phase delays caused by the presence of the phase glitches, $\sigma_{\phi_{gli}}$, will be treated as a systematic errors. Therefore, for each pulse phase delay value we computed the uncertainty as $\sigma_{\Delta\phi}=(\sigma^2_{\phi_{stat}}+\sigma^2_{\phi_{gli}})^{1/2}$, where $\sigma_{\phi_{stat}}$ is the statistical error associated to each of the pulse phase delays estimated.

\begin{table*}

\begin{tabular}{l | c c c}
Parameters             & Outburst 1996 & GBM project & Outburst 2014 \\
\hline
\hline
Orbital period $P_{orb}$ (days) & 11.8337(13) & 11.836397&11.8358(5)\\
Projected semi-major axis a sin\textit{i/c} (lt-s) & 2.6324(12) &2.637 & 2.639(1) \\
Ascending node passage $T^{\star}$ (MJD) & 50076.6968(18)& 56692.73970& 56692.739(2)\\
Eccentricity  E &$ < 1.1 \times 10^{-3} $& 0.0000 & $< 6 \times 10^{-3}$\\
Spin frequency $\nu_0$ (Hz) &2.141004032(14)&2.1411117 & 2.141115281(8)\\
Spin frequency derivative $\dot{\nu}_0$ (Hz s$^{-1}$) & 9.228(27)$\times10^{-12}$ & - &$^*$1.644(5)$\times10^{-12}$ \\
Epoch of $\nu_0$ and $\dot{\nu}_0$, $T_0$ (MJD) & 50085.0&56677.0& 56693.8 \\
\hline
\hline
\end{tabular}
\caption{Orbital parameters of \gro{} obtained by analysing the 1996 outburst \citep[first column;][]{Finger96}, the combination of 1996 and 1997 outbursts combined with the 2014 data collected and analysed by \fermigbm{} project (second column; \url{http://gammaray.nsstc.nasa.gov/gbm/science/pulsars}), and the 2014 outburst investigate in this work (third column). Errors are at 1$\sigma$ confident level, while no errors are available for the \fermigbm{} project. *The value of $\dot{\nu}_0$ reported for the 2014 outburst refers to the accretion torque modelling described in Eq.~\ref{eq:nu_dot_start}, hence it represents the local value of $\dot{\nu}$ at the beginning of the outburst.}
\label{tab:par2}
\end{table*}

\subsubsection{Accretion torque}
\label{sec:acctor}
To investigate the spin evolution of the source during the outburst, we track the evolution of the pulse phase delays ($\Delta \phi(t)$) as function of time. As explained in \citet[][see also references therein]{Burderi07}, the main idea is based on few simple assumptions that we summarise as follows:

\begin{enumerate}

\item matter accretes through a Keplerian disc truncated at the magnetospheric radius, $R_m$, because of the interaction with the (dipolar) magnetic field of the NS. At this radius the accreting matter is forced to corotate with the magnetic field and it is funnelled toward the magnetic poles causing the pulsed emission. The magnetospheric radius is commonly related to the radius at which the magnetic energy density equals energy density of the accreting matter assumed in free fall (also known as Alfv\'en radius, $R_{\rm A}$), via the relation\footnote{See \citet{Bozzo2009a} for the uncertainties and a summary of the assumptions behind this definition.}:
\begin{equation}
R_{\rm m} = \xi R_{\rm A} = \xi (2GM)^{-1/7} \mu^{4/7} \dot{M}^{-2/7},
\label{eq:mag_rad}
\end{equation}
where $\xi$ is a model-dependent dimensionless number usually between 0 and 1 \citep{Ghosh79b, Wang96, Burderi98a}, $G$ is the gravitational constant, $M$ the NS mass, $\mu$ denotes the star's magnetic dipole moment, and $\dot{M}$ is the mass accretion rate;
\item matter accreting on the NS carries its specific Keplerian angular momentum at the magnetospheric radius, $\ell = (GM R_{\rm m})^{1/2} $. Therefore, a material torque $\tau_{\dot{M}} = \ell \dot{M}$ is exerted onto the NS;
\item mass accretion rate $\dot{M}$ can be well traced by the bolometric luminosity ($L$) via the relation $\dot{M}= L/\eta (G M/R)$, where $\eta \lesssim 1$ is a conversion factor which indicates the efficiency of the accretion process in units of standard accretion efficiency onto NS \citep[$GM\dot{M}/R$; e.g.][]{Frank02}, where $R$ is the NS radius;
\item we considered the material torque $\tau_{\dot{M}}$ as the only torque acting on the NS. Any form of threading of the accretion disc by the magnetic field has been ignored \citep[see, e.g.,][for a detailed description of magnetic threading models]{Ghosh79a,Ghosh79b,Wang87, Wang95,Wang96, Wang97a, Rappaport04}.
\end{enumerate}  
\begin{figure*}
\centering
\includegraphics[width=1\textwidth]{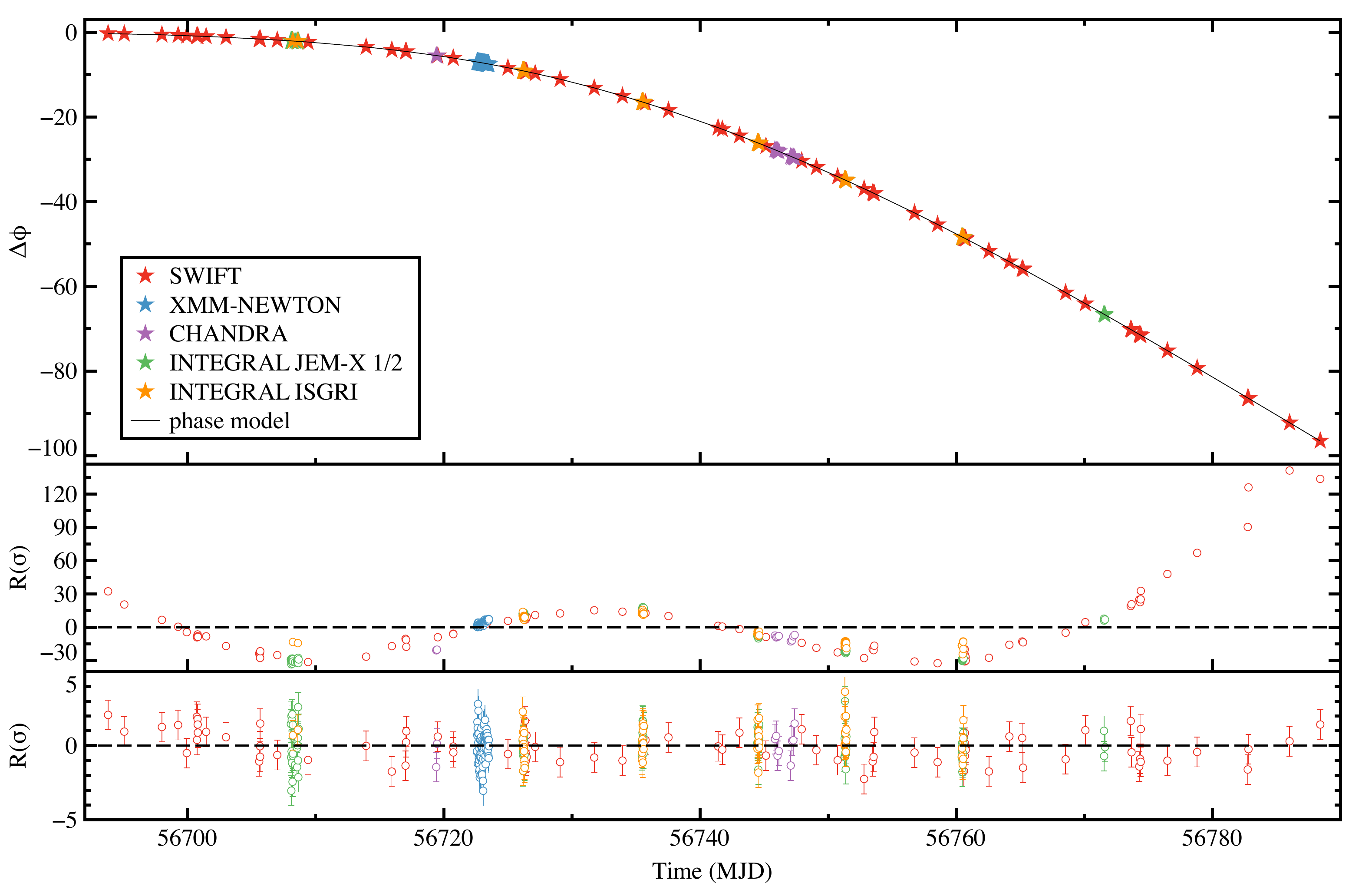}
\caption{\textit{Top panel -} Pulse phase delays as a function of time computed by epoch-folding \swiftxrt{} (red), \xmm{} (blue), \chandra{} (purple), \intejx{} \textit{1/2}, and \inteis{} data at the spin frequency $\nu_0 = 2.141115128$ Hz, together with the best-fit model (black line, see text). \textit{Middle panel -} Residuals in units of sigma with respect to a model that assumes a constant $\dot{M}$. \textit{Bottom panel -} Residuals in units of sigma with respect to the best fitting model.}
\label{fig:fit_fasi}
\end{figure*}
Adopting the above mentioned assumptions, the spin frequency derivative $\dot{\nu}$ can be written as follows:

\begin{equation}
\begin{split}
\label{eq:tq}
\dot{\nu} &= \frac{\dot{\Omega}}{2 \pi}=\frac{\tau_{\dot{M}} }{2 \pi I} = \frac{\ell R L}{2 \pi I \eta GM}=\\
&=2^{-15/14}\,\xi^{1/2}\,\mu^{2/7}\,(GM)^{-3/7}\,(I\pi)^{-1}\,R^{6/7}\,\eta^{-6/7}\,L^{6/7},
\end{split}
\end{equation}
\noindent
where $I$ is the moment of inertia of the NS. The aforementioned expression is correct if the variation of $I$ during the accretion process is negligible. In the follow, we quickly demonstrate that the previous assumption is correct for the 2014 outburst of \gro{}. From definition the accretion torque is expressed as:

\begin{equation}
\label{eq:idot}
\tau=\frac{d\left(I\omega\right)}{dt}=I\dot{\omega}+\omega\dot{I},
\end{equation}
where $\omega=2\pi\nu$ is the angular velocity of the NS. Expressing the moment of inertia of the NS as $I\propto MR^2$, and generalising the NS mass--radius relation as $R\propto M^{\rho}$ with $\rho=-1/3$,  $\rho\sim0$,  and  $\rho=+1/3$ for pure degenerate non-relativistic neutron gas, realistic NS equation of states, and incompressible ordinary matter, respectively. Eq.~\ref{eq:idot} can then be written as:
\begin{equation}
\label{eq:idot2}
\tau=I\dot{\omega}\left[1+\frac{\dot{M}\nu}{M\dot{\nu}}(1+2\rho)\right].
\end{equation}
We consider, for the 2014 outburst of \gro{}, an average $\dot{M}$ of $\sim3\times10^{-9}M_{\odot}/yr$, $\dot{\nu}\sim2\times10^{-12}$ Hz/s, and $\rho\sim0$, we find $\tau=I\dot{\omega}\left[1+1\times10^{-4}\right]\simeq I\dot{\omega}$, in agreement with our starting assumption.

As clearly shown in Eq.~\ref{eq:tq}, the spin frequency derivative caused by the accretion torque depends upon the luminosity as $\dot{\nu}\propto L^{6/7}$. Moreover, since all the quantities reported in Eq.~\ref{eq:tq} are basically constant during the outburst, except for the luminosity, the variation of the spin frequency derivative as a function of time will depend upon the evolution of the luminosity, as $\dot{\nu}(t)\propto L^{6/7}(t)$. To verify this model we write the former relation assuming a dependency of  the luminosity on a generic power, $\beta$. We can then rewrite Eq.~\ref{eq:tq} as a function of time like:
\begin{equation}
\label{eq:nu_dot_L}
\dot{\nu}(t)= \dot{\nu}(T_1) \left[\frac{L(t)}{L_1}\right]^{\beta},
\end{equation}
where
\begin{equation}
\label{eq:nu_dot_start}
\dot{\nu}(T_1)= 2^{-15/14}\,\xi^{1/2}\,\mu^{2/7}\,(GM)^{-3/7}\,(I\pi)^{-1}\,R^{6/7}\,\eta^{-6/7}\,L_1^{\beta},
\end{equation}

\noindent
with $\dot{\nu}(T_1)$ and $L_1$ corresponding to the frequency derivative and the bolometric luminosity at the beginning of the dataset considered in this work, respectively. 

To derive the luminosity for each spectrum we assumed the unabsorbed flux estimated in the energy range 0.1-100 keV to be a good proxy of the bolometric luminosity. We then express the fluxes reported in Fig.~\ref{fig:flux} as a suitable combination of linear functions of time. We divided the entire outburst into a series of $N$ intervals ($n=1,...,N$), defined by their starting times $T_{n}$ at which a flux value has been determined. The flux at the beginning of the first interval ({\it i.e.} for $t = T_{1}$) is  $F_{1}$. Assuming that the flux in each interval, $F_n(t)$, increases or decreases linearly at a certain rate $k_n$. Hence, we can represent the fluxes versus time analytically as follows:
\begin{equation}
\begin{array}{ll}
\label{eq:L_vs_t}
F_{1}(t) = & F_{1}\times [1 +  k_{1} (t-T_{1})] \\
F_{2}(t) = & F_{2}\times [1 +  k_{2} (t-T_{2})]  \\
...       \,\,\,\,\,\,\,\,\, = & ... \\
F_{N}(t) = & F_{N}\times[1 +  k_{N} (t-T_{N})]. \\
\end{array}
\end{equation}
Combining Eq.~\ref{eq:nu_dot_L} and Eq.~\ref{eq:L_vs_t} we find the spin frequency derivative caused by the material torque in each time interval ($n=1,...,N$) to be:
\begin{equation}
\label{eq:nudot4} 
\dot{\nu}_{n}(t) = \dot{\nu}(T_1) \times \left\{\frac{F_{n}}{F_{1}}\times [1 +  k_{n} (t-T_{n})]\right\}^{\beta}. \\
\end{equation} \\
As discussed in Sec.~\ref{sec:timing}, if the folding frequency $\nu_0$ is close to the real spin frequency of the source and the spin frequency variation throughout the outburst is small, $\Delta \phi(t)$ can be expressed as
\begin{eqnarray}
\Delta \phi(t)&=&\phi_0-[\nu(t)-\nu_0]\times(t-T_0)\nonumber\\
&-&\int_{t_0}^tdt'\int_{T_0}^{t'}\dot{\nu}(t'')dt''+R_{orb}(t)
\end{eqnarray}
where $R_{orb}(t)$ represents the Roemer delay related to the orbital parameters differential corrections \citep[see e.g.,][]{Deeter81}. For the generic $n-$th time interval we can write the phase delay with respect to the folding frequency $\nu_0$ at the reference epoch $T_0$ as
\begin{eqnarray}
\label{eq:phin}
\Delta\phi_{n}(t) &= &
\Delta\phi_{n-1}(t=T_{n}) - [\nu(t)-\nu_0] \times (t - T_{n})+\\
 &-& \dot{\nu}(T_1) \times \Biggl\{\left[ \sum_{i=1}^{n-1} \left(\frac{F_i}{F_1}\right)^{\beta}\times I_1(\beta, i, T_{i+1}) \right] \times (t - T_{n}) + \nonumber\\
 &+& \left(\frac{F_n}{F_1}\right)^{\beta} \times I_2(\beta, n, t)\Biggl\}+R_{orb}(t)\nonumber\\
\nonumber
\end{eqnarray}
\noindent
where $I_1(\beta,n, T_{n+1}) =  \int_{T_{n}}^{T_{n+1}} [1 +  k_{n} (t'-T_{n})]^{\beta} \hspace{0.1cm} dt'$ and $I_2(\beta,n, t) = \int_{T_{n}}^{t} I_1(\beta,n, t')  \hspace{0.1cm} dt'$.
Using Eq.~\ref{eq:phin}, we then fitted the evolution of the observed pulse phase delays as a function of the parameters $\phi_0$, $\Delta\nu(T_0)$, $\dot{\nu}(T_1)$, $\beta$ and the correction to the orbital parameters. Best-fit parameters are shown in Table~\ref{tab:par1}. In Fig.~\ref{fig:fit_fasi} (bottom panel) we report the residuals in units of $\sigma$ with respect to the best-fit model described in Eq.~\ref{eq:phin}. The value of $\tilde{\chi}^2\sim 1.12$ (with 313 d.o.f.) combined with the distribution of the residuals around zero (Fig.~\ref{fig:fit_fasi}, bottom panel), clearly shows that the adopted model well describes the pulse phase delays evolution during the outburst. We tested the validity of these results by replacing the description of the flux as a function of time reported in Eq.~\ref{eq:L_vs_t} with a numerical integration of the flux function obtained by interpolating each pair of flux measurements with a cubic polynomial (cubic spline). We found no discrepancy, within errors, between the two methods.

\noindent
Finally, although the the reduced $\chi^2$ obtained by fitting the pulse phases with the accretion torque model alone is reasonably good (1.06 with 313 d.o.f.), we tried to fit the pulse phase delays by adding a quadratic component to investigate the presence of spin-down activity. This new model well fits the data leading to a value of $\tilde{\chi}^2\sim1.03$ for 312 d.o.f. By including the quadratic component we register a $\Delta\chi^2\sim10$ for a decrease in degrees of freedom of one, which corresponds to a F-test probability of 0.0021 ($\sim3\sigma$). Given the low level of significance of this component we decided not to included it in the final model. 
\begin{table}
\begin{center}
\begin{tabular}{l| c | c}
Fit parameters        &     outburst 2014 & MCM  \\
\hline
\hline
$\phi_0$&-0.283(8) & -0.29(10) \\
$\nu(T_0)$ (Hz)& 2.141115281(9) & 2.141115273(10)\\
$\dot{\nu}(T_1)$ &1.644(5)$\times 10^{-12}$ & 1.648(52)$\times 10^{-12}$\\
a sin\textit{i/c} (lt-s) & 2.639(1) & 2.639(3)\\
P\textit{$_{orb}$} (s) &11.8358(5) & 11.8358(14) \\
T$^*$ (MJD) &  56692.739(2) &  56692.739(5)\\
e & $< 6 \times 10^{-3}$&$< 1.4 \times 10^{-2}$ \\
$\beta$ & 0.932(7) & 0.960(30)\\

\hline
$\chi^2$/d.o.f. & 331.8/313 & --
\end{tabular}
\caption{First column shows the best fit parameters of \gro{} obtained by modelling with Eq.~\ref{eq:phin} the pulse phase delays calculated by folding the \swift{}, \inte{}, \chandra{}, and \xmm{} observations of the source. In the second column are reported the parameters of the torque model as the result of the Monte Carlo simulations (MCM, see Section~\ref{sec:discussion} for more details). Uncertainties are given at 1$\sigma$ confidence level.}
\label{tab:par1}
\end{center}
\end{table}

\section{discussion}
\label{sec:discussion}
In this work we investigated the temporal evolution of the coherent X-ray pulsations shown by \gro{} during its 2014 outburst, based on the whole available dataset collected by \swiftxrt{}, \swiftbat{}, \xmm{}, \inte{} (ISGRI and JEM-X 1/2), and \chandra{}. 

\noindent
The results obtained from the timing analysis clearly show a correlation between the NS angular acceleration ($\dot{\nu}$) and the amount of energy that it releases per unit of time in the X-rays ($L$), corroborating the hypothesis that the bolometric X-ray luminosity is a good tracer of the amount of matter that accretes onto the NS surface. Another hint for the relation $\dot{\nu}(t) \propto \dot{M}(t)$ can be seen in the middle panel of Fig.~\ref{fig:fit_fasi} where we reported the residuals in units of $\sigma$ of the pulse phase delays with respect to a model with constant $\dot{M}$. From the residuals we can clearly conclude that the pulse phase delays are not compatible with such a model. Furthermore, it is worth noting that the material torque model used to fit the pulse phase evolution during the 2014 outburst of the source is able alone to well describe the data (with a reduced $\chi^2\sim1.06$ for 313 d.o.f.). This strongly suggests that, at least for this outburst, the spin-down component is almost absent, probably because the material torque overcomes any magnetic threading.  

To further investigate in detail the accretion torque mechanism, short-term torque measurements are required. Basically all such models predict that the magnetospheric radius should decrease as $\dot{M}$ increases. Ignoring any kind of threading mechanism between the accretion disc matter and the magnetic field, standard accretion disc theories \citep[e.g.,][]{Ghosh79a} predict $R_{\rm m} \propto \dot{M}^{-2/7}$, implying that the X-ray pulsar should accelerate at a rate $\dot{\nu} \propto \dot{M}^{6/7}$. Depending on the prescription for the interaction between magnetic field and accreting matter at the truncation radius, it is possible to find slightly different relations between $\dot{\nu}$ and $\dot{M}$, such as $\dot{\nu} \propto \dot{M}^{9/10}$ \citep{Kluzniak07,Kulkarni13}, $\dot{\nu} \propto \dot{M}^{0.87}$ \citep[model 1G--GPD;][]{Ghosh96}, $\dot{\nu} \propto \dot{M}^{0.92}$ \citep[model 1R--RPD;][]{Ghosh96}, $\dot{\nu} \propto \dot{M}^{0.76}$ \citep[model 2B--GPD;][]{Ghosh96}, and $\dot{\nu} \propto \dot{M}^{0.15}$ \citep[model 2S--GPD;][]{Ghosh96}. These predictions can be tested by measuring the correlation between the time evolution of the pulse phases (torque) and the bolometric luminosity, generally considered a good proxy of the mass accretion rate. 

As reported in Table~\ref{tab:par1}, from the fit of the latest outburst, \gro{} does not seem to be in agreement with the simple $\dot{\nu} \propto \dot{M}^{6/7}$ prescription since we find $\beta=0.932\pm0.007$. A similar work for \gro{} has been done by \citet{Bildsten97b} using data collected with BATSE. They found $\beta = 0.957\pm0.026$ which is interestingly similar to the value we find combining data from \swift{}, \inte{}, \chandra{}, and \xmm{}. Other sources has been explored, such as A0535+26 \citep[$\beta= 0.951\pm0.026$, see][for more details]{Bildsten97b}, EXO 2030+375 \citep[$\beta= 1.21\pm0.13$, see][]{Parmar1989a} and IGR J17480--2446 \citep[$\beta= 1.07\pm0.03$, see][]{Papitto12}. We note that similarly to \gro{}, also for A0535+26 has been reported a measurement of the parameter $\beta$ almost $4\sigma$ apart from the assumed $\beta=6/7$. On the contrary, the values reported for EXO 2030+375 as well as for IGR J17480--2446 are compatible within errors ($3\sigma$ confidence level) with $\beta=6/7$.

In an incomplete sampling of the outburst, flux changes with time between consecutive observations can add to the systematic
uncertainties in the timing solution. To quantify this, we created a thousand flux curves starting from the one reported in Fig.~\ref{fig:flux}, and proceeding as follows: i) for each flux measurement reported in Fig.~\ref{fig:flux} we generated a random value assuming a normal distribution with mean parameter and standard deviation equal to the corresponding flux measurement and its $1\sigma$ uncertainty, respectively ii) for each pair of consecutive flux measurements reported in Fig.~\ref{fig:flux} we randomly choose one of the three possible interpolation methods: a) linear interpolation (the same used to implement the torque model in Section~\ref{sec:timing}); b) piecewise constant interpolation with value corresponding to the first flux measurement of the interval considered; c) piecewise constant interpolation with value corresponding to the second flux measurement of the interval considered. Each of the simulated flux curves has then been then used to fit the pulse phase delays reported in the top panel of Fig.~\ref{fig:fit_fasi}. 
\begin{figure}
\centering
\includegraphics[width=0.48\textwidth]{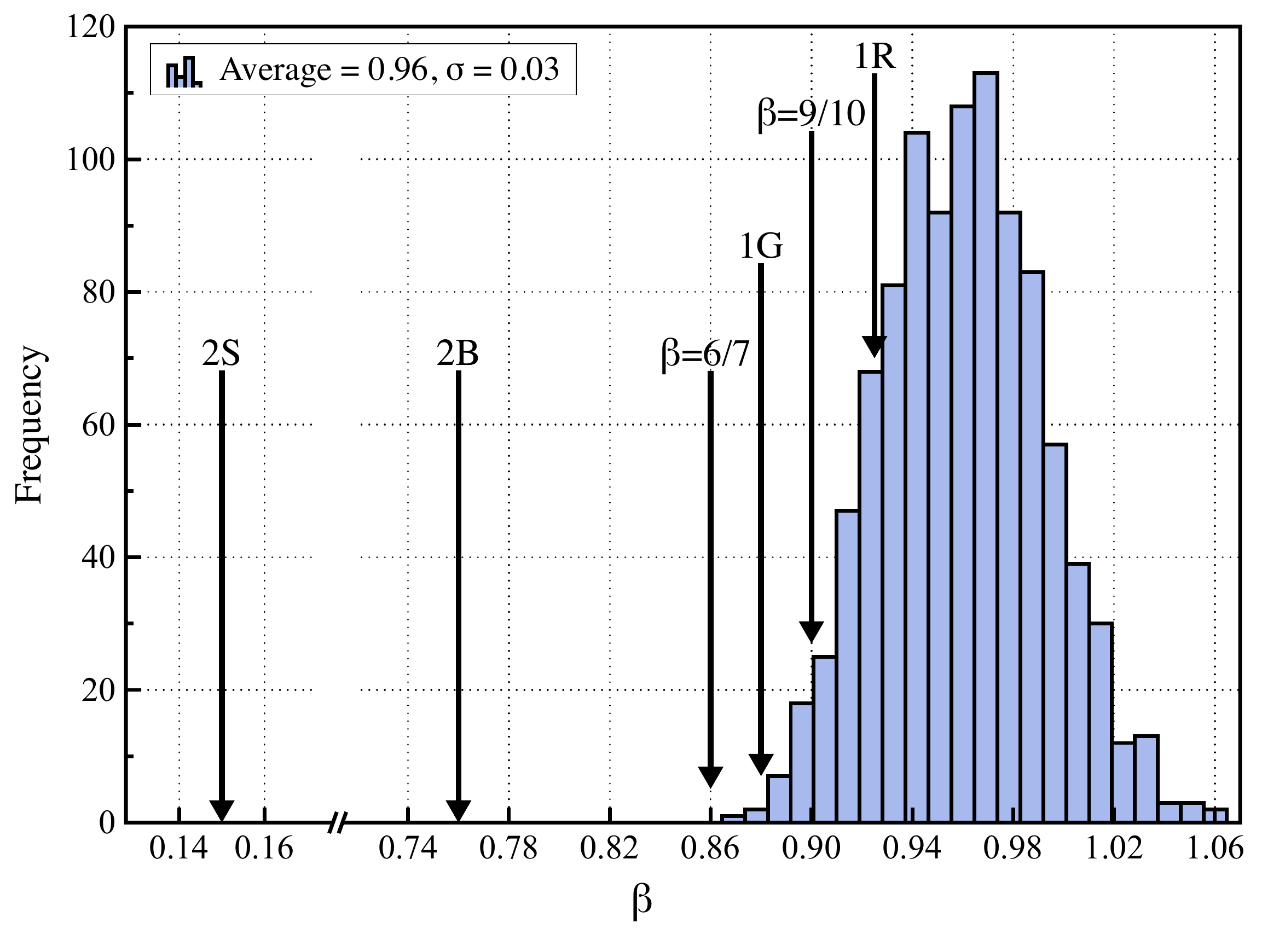}
\caption{Distribution of the parameter $\beta$ from the accretion torque model obtained from the Monte Carlo analysis with 1000 simulations. The distribution has an average value $\beta = 0.96$ and a standard deviation $\sigma = 0.03$ (see Section~\ref{sec:discussion} for more details). Arrows represent values of $\beta$ associated with different theoretical models, see text for references.}
\label{fig:histo}
\end{figure}
In Fig.~\ref{fig:histo} we show the distribution of the parameter $\beta$ as a result of the simulations described above. As expected the mean of the distribution $\bar{\beta}_{MCM} = 0.96$ is very similar to the best-fit parameter reported in Table~\ref{tab:par1}, on the other hand the standard deviation $\sigma_{MCM} = 0.03$ differs more than a factor of 4 in magnitude compared to the $\sigma_{\beta}$ statistical error from the fit. From this result we can define a more realistic range of the parameters $\beta$ between 0.93 and 1 (1$\sigma$ interval). Moreover, in Fig.~\ref{fig:histo} we reported for reference the values of $\beta=6/7$ predicted by \citet{Ghosh79a} from which our result differs $\sim3\sigma$. Our result is also consistent within $2\sigma$ with the models proposed by \citet{Kluzniak07, Kulkarni13} and by \citet[model 1R--RPD;][]{Ghosh96}. Models such as 2B and 2S \citep{Ghosh96} describing a two-temperature optically thin mass-pressured-dominated disc, are clearly not compatible with our findings. We emphasise that the combination of the flux uncertainties and the sampling of the outburst do not give us, for this source, the sensitivity to investigate small differences parametrised by $\beta$ and predicted by the different models. Nonetheless, we notice that the diamagnetic disc assumption is a simplification of the real disc magnetosphere interaction, which is probably better described by the so called threaded disc model \citep[see e.g.,][]{Ghosh79a,Ghosh79b,Wang87, Wang95,Wang96, Wang97a, Rappaport04, Bozzo2009a}. Nonetheless, as shown by \citet{Bozzo2009a}, threaded disc models prescribe more complex and variable behaviours of the magnetospheric radius as a function of luminosity depending on the theoretical prescriptions adopted to describe the matter-magnetic field interaction. An implementation of the torque in the framework of the threaded disc model is beyond the scope of this work and it will be discuss elsewhere. In the last section of this work we try to investigate the presence of a spin-down component from the analysis of the pulse phase delays and we discuss the results suggesting a possible link with threaded disc models.

Finally, combining the frequency derivative and the flux estimation we can constrain the source distance. To do that we followed the prescription adopted by \citet{Burderi98a}, with the following assumptions:
\begin{itemize}
\item magnetic dipole perfectly orthogonal with respect to the disc plane; 
\item perfectly diamagnetic accretion disc, which implies a dipolar magnetic field completely screened by sheet currents at the disc truncation radius. Consequently, the magnetic pressure exerted on the innermost layer of the disc (where the sheet currents flow), can be written as: 
\begin{equation}
P_{mag} = \frac{1}{8\pi} \left[B_{in}^{2}-B_{out}^{2}\right]
\end{equation}
\cite[see e.g.,][]{Purcell}, where B$_{in}$ and B$_{out}$ are the inner and outer magnetic fields infinitesimally close to the sheet currents.  The perfectly diamagnetic condition implies B$_{in}$=2B$_{ext}$ and B$_{out}$=0, where B$_{ext}$ is the external field. Adopting spherical coordinates, the component along the z-axis of a magnetic field produced by a magnetic dipole of intensity $\mu$ placed at the origin can be written as B$_{z}$=$\mu[3\cos^2 \theta-1]/r^3$. For $\theta=90^\circ$, B$_{z}(\theta)=B_{eq}=\mu/r^3$, the magnetic pressure at the equator is
\begin{equation}
P_{mag} = \frac{\mu^2}{2\pi r^6}
\end{equation}
\item standard Shakura--Sunyaev disc pressure for gas-pressure-dominated regions and free-free opacity (zone C) \cite[see e.g.,][]{Frank02}, that can be expressed as:
\begin{equation}
P_{dis} = 2.8\times10^5 \kappa^{-1}_{0.615}\, \alpha^{-9/10}\,\dot{M}_{{-9}}^{17/20}\,m^{7/8}\,R^{-21/8}_{10} dyne\ cm^{-2},
\end{equation}
where $\kappa_{0.615}$ represents the mean molecular weight in units of 0.615 adopted for fully ionised cosmic mixture of gases \cite[see e.g.,][]{Frank02}, $\alpha$ parametrises the disc viscosity in the standard Shakura--Sunyaev solution, $\dot{M}_{{-9}}$ is the mass accretion rate in units of $10^{-9}\,{M_{\odot}}$ per year, m is the NS mass in units of M$_{\odot}$, and R$_{10}$ is the distance from the NS in units of $10^{10}$ cm. 
\end{itemize}

Equating the magnetic and disc pressures and combining the Alfv\'en radius expression reported on Eq.~\ref{eq:mag_rad}, we can derive the parameter $\xi$ as:
\begin{equation}
\xi = 0.315\, \kappa^{8/27}_{0.615}\,\alpha^{4/15}\, \mu_{26}^{4/189}\,\dot{M}_{{-9}}^{32/945}\,m^{76/189},
\label{eq:xi}
\end{equation}
where $\mu_{26} = B_{8}\,R^3_6 = 10^{26}$ G\,cm$^{3}$ is the NS magnetic moment assuming a magnetic field B of $10^{8}$ Gauss and a NS radius of $10^6$ cm. Expressing the mass accretion rate as a function of the observed flux and combing the material accretion torque (Eq.~\ref{eq:nu_dot_start}) with the magnetospheric radius expression (Eq.~\ref{eq:mag_rad}) and the parameter $\xi$ (Eq.~\ref{eq:xi}), we can express the distance $d$ of the source as follows:
\begin{multline}
\label{eq:dist}
d=2.34\left(\frac{\eta}{R_6\,F_{-8}}\right)^{1/2}\\
\left[\alpha^{-9/2}_{0.5}\,(I_{45}\,\dot{\nu}_{-12})^{135/4}\,\mu^{-10}_{30}\,m^{33/4}\,\kappa^{-5}_{0.615}\right]^{1/59}\,\text{kpc}
\end{multline}
where $\alpha_{0.5}$ is the disc viscosity in units of 0.5, $I_{45}$ is the moment of inertia of the NS in units of $10^{45}$ g\,cm$^2$, $\dot{\nu}_{-12}$ is the frequency derivative of the NS in units of $10^{-12}$ s$^{-2}$, $\mu_{30} = B_{-11}\,R^3_6 = 10^{30}$ G\,cm$^{3}$ is the NS magnetic moment assuming a magnetic field B of $10^{11}$ Gauss, and F$_{-8}$ is the unabsorbed flux (in the energy range 0.1--100 keV ) of the source in units of $10^{-8}$ erg s$^{-1}$ cm$^{-2}$. In the following we adopt a NS mass $m=1.4$ M$_{\odot}$, and an efficiency conversion factor $\eta=1$. Considering the superficial magnetic field $B=5.27\times10^{11}$ G inferred by \citet{DAi15} from the detection of the cyclotron absorption line at 4.7 keV, we can estimate the NS magnetic moment $\mu_{30}=0.527$. From the best-fit value of spin frequency derivative $\dot{\nu}(T_1)$ and the corresponding flux value $F_{1}=9.94\times10^{-9}$ erg s$^{-1}$ cm$^{-2}$ we estimate a source distance of $d\sim3.6$ kpc for $\alpha=0.5$. 

\begin{figure}
\centering
\includegraphics[width=0.48\textwidth]{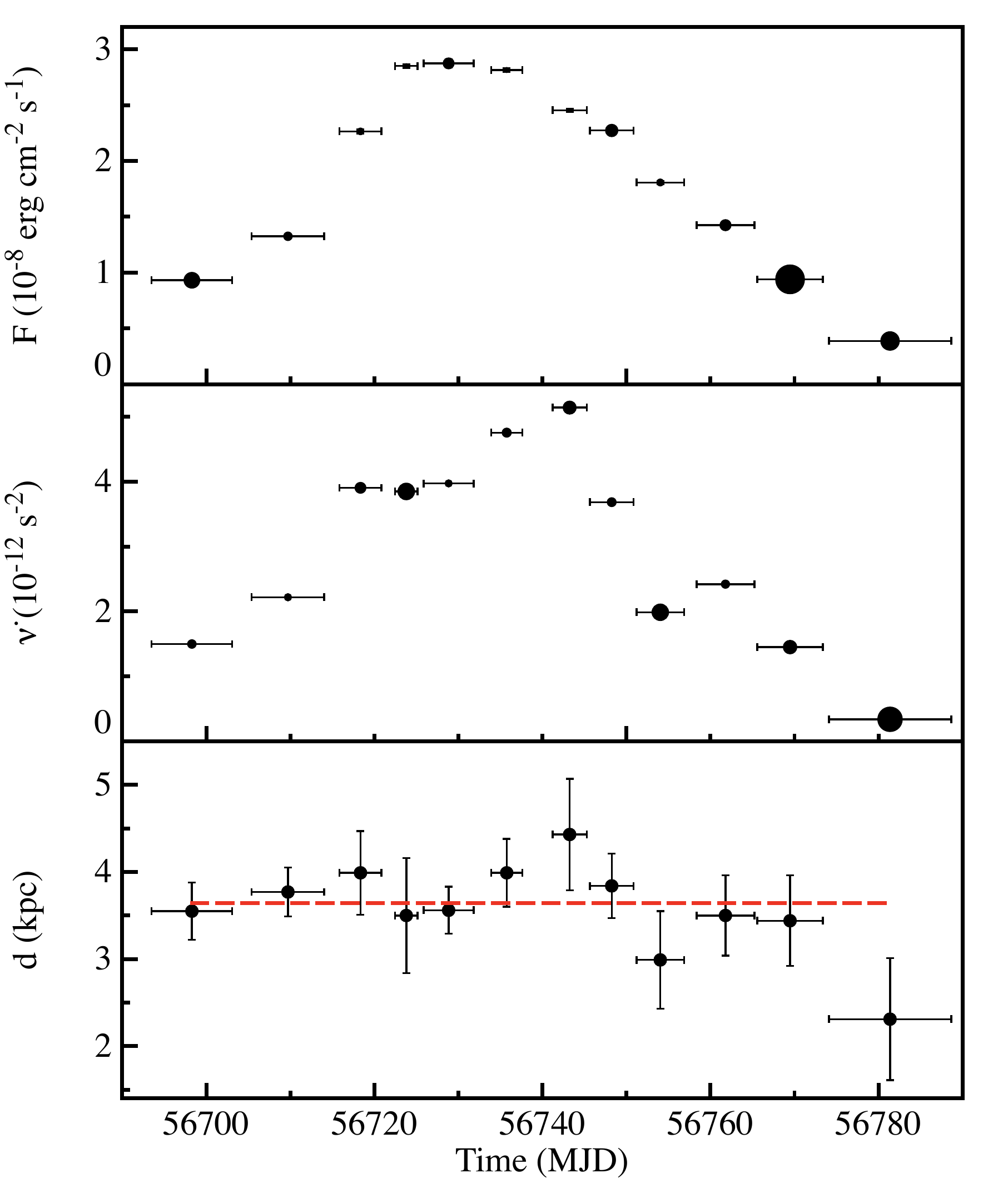}
\caption{\textit{Top panel - } Unabsorbed flux estimated in the energy range 0.1--100 keV sub-diving the Fig.~\ref{fig:flux} in non-overlapping intervals of different length (see Section~\ref{sec:discussion} for more details). \textit{Middle panel - } Spin frequency derivative calculated for each interval by fitting the pulse phase delays with a quadratic function. The size of the symbols reported in the \textit{top} and \textit{middle} panel are proportional to the flux and spin frequency derivative relative errors, respectively. \textit{Bottom panel - } Source distance estimates calculated using Eq.~\ref{eq:dist} for each data interval (black points), and associated weighted mean value (red dashed line).}
\label{fig:distance}
\end{figure}

To verify whether the distance estimation suffers from the modelling of the accretion torque applied for the timing analysis, we performed the following test: i) we divided the outburst in smaller non-overlapping time intervals with variable length depending on their statistics; ii) for each interval we fitted the corresponding pulse phase delays with a quadratic function using as a reference time the centre of the interval, furthermore we fitted the corresponding flux curve with a linear function; iii) finally using Eq.~\ref{eq:dist} we estimated the source distance combining the estimates of $\dot{\nu}$ and $F$ for each interval and the magnetic field from \citet{DAi15}. In Fig.~\ref{fig:distance}, top and middle panels, we report the X-ray flux and the spin frequency derivative for each of the intervals we selected. It is interesting to note how the two quantities show a very similar time evolution, which is in agreement with the result obtained from the timing analysis. On the bottom panel we show the associated values of the source distance (black points), and the corresponding weighted mean (red dashed line) at the value $d=3.7(5)$ kpc which is consistent with the distance estimated from the fit of the whole outburst.     

\begin{figure}
\centering
\includegraphics[width=0.48\textwidth]{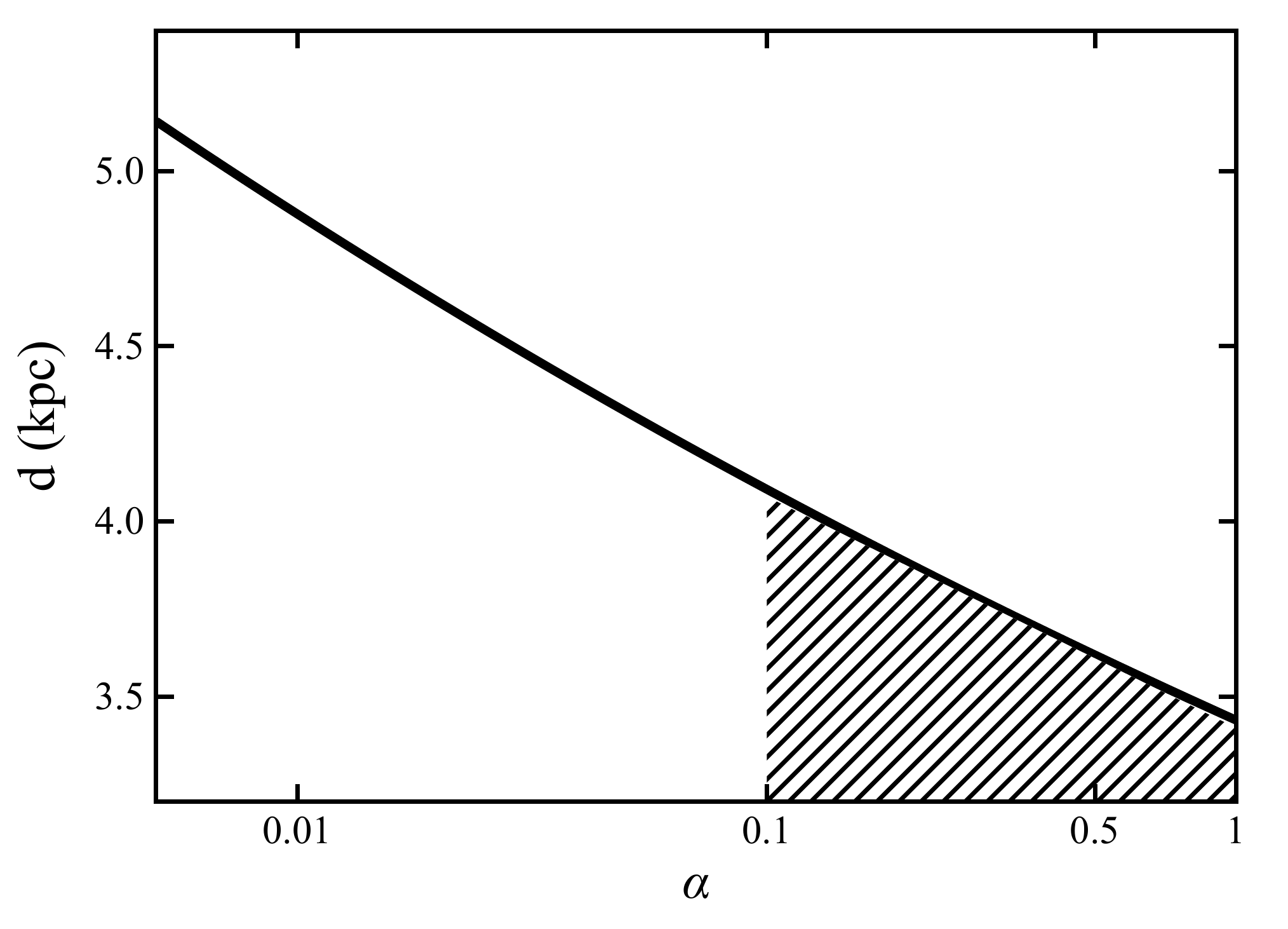}
\caption{Source distance estimates in kpc as a function of the disc viscosity $\alpha$, calculated combing Eq.~\ref{eq:dist} with the assumption that the truncation disc radius is within the NS radius and the co-rotation radius $R_{co}$. The shaded region highlights the range of the distance values estimated for viscosity values between 0.1 and 1.}
\label{fig:dis-phi}
\end{figure}

As previously mentioned, we estimated the distance assuming a value of $\alpha=0.5$. For completeness we calculated the source distance within a plausible range of $\alpha$ values. We also imposed the condition $R_{\rm{NS}} < R_m \lesssim R_{co}$, where $R_{co}=(GM/\omega^2_{s})^{1/3}$ is the co-rotation radius. In Fig.~\ref{fig:dis-phi} we report the source distance in kpc as a function of the parameter $\alpha$. We note that at $\alpha\sim1$ corresponds $R_m \gtrsim R_{co}$ for which the source should switch to the propeller regime \citep[][]{Illarionov75, Ghosh79a, Wang87, Rappaport04}. On the other hand, we assumed the lowest value of $\alpha$ to be $\sim0.01$ \citep[see][and references therein]{King2007a}. Converting the previous finding in terms of the parameter $\xi$, we find that at least for GRO J1744--29 $\xi$ is limited in the range 0.13--0.46 that does not include neither the value $\xi=0.5$, although very close, prescribed by \citet{Ghosh79a}, nor $\xi\sim1$ predicted by \citet{Wang96}. Interestingly the reported $\xi$ range includes the value ($\xi\sim0.2$) inferred by \citet{DAi15} from the spectral analysis of the latest outburst of \gro{}.  As shown in Fig.~\ref{fig:dis-phi}, we estimated the source distance to range between $\sim3.4$ kpc and $\sim5.1$ kpc. The range of distances can be further reduced making some considerations on typical values of $\alpha$ for LMXBs. As reported by \citet{King2007a}, the viscosity parameters for these type of objects ranges between 0.1--1, if that is the case then we can constrain the distance between 3.4 kpc and 4.1 kpc (dashed area in Fig.~\ref{fig:dis-phi}). 

The range of values estimated from our timing analysis of the 2014 outburst of \gro{} is consistent with the source distance estimates reported in literature. Based on the high values of the equivalent hydrogen column $N_H\sim(0.5-1)\times10^{23}$ atoms cm$^{-2}$, several authors placed an upper limit on the source distance in the range 7.5--8.5 kpc in correspondence with the Galactic center \citep[e.g., ][]{Giles96,Augusteijn97, Nishiuchi99}. On the other hand, a lower limit of 1.5 kpc and 2 kpc have been set from the analysis of the optical and near-infrared possible counterparts of the source, respectively \citep[e.g., ][]{Cole97}. Consistent results were also reported independently by \citet{Gosling07} and \citet{Wang07}, that from the photometry and spectroscopy of the near-infrared counterpart of \gro{} estimated a distance of $3.7\pm1$ kpc and $\sim4$ kpc, respectively. It is worth mentioning that the last two distance values are interestingly compatible with our estimation, $d\sim3.6$ kpc, calculated assuming $\alpha=0.5$. If indeed \gro{} is located at approximately 4 kpc,  we should revise the luminosity estimates from super-Eddington ($\sim 5$L$_{Edd}$) to nearly Eddington in the 1996 outburst, and from Eddington to half-Eddington for the 2014 outburst. 

Nonetheless, the reported spin-up evolution as well as the presence of marginally significant spin-down component make \gro{} a good candidate to further investigate threaded disc models. However, as highlighted in this work, to achieve the level of accuracy required to disentangle the threaded effects high statistic data combined with an accurately sampled outburst are required, both of them hopefully achievable with future X-ray missions.
On the other hand, we stress that even if expected from theoretical considerations and observations of spin derivative reversal in other sources \citep[e.g., GX 1+4 ][]{Gonzalez-Galan2012a}, the data analysed in this work did not allow a clear detection of a spin-down component. In our opinion, this proves that, at least during the bright phase of the 2014 outburst discussed in this work, the spin-down component is virtually negligible likely reflecting the fact that the material torque overcomes any magnetic threading.

\section*{Acknowledgments}
A. R. gratefully acknowledges the Sardinia Regional Government for the financial support (P. O. R. Sardegna F.S.E. Operational Programme of the Autonomous Region of Sardinia, European Social Fund 2007-2013 - Axis IV Human Resources, Objective l.3, Line of Activity l.3.1). P. E. acknowledges funding in the framework of the NWO Vidi award A.2320.0076. This work was partially supported by the Regione Autonoma della Sardegna through POR-FSE Sardegna 2007-2013, L.R. 7/2007, Progetti di Ricerca di Base e Orientata, Project N. CRP-60529. We acknowledge a financial contribution from the agreement ASI-INAF I/037/12/0.

\section*{Appendix: pulse phase systematic uncertainties on the accretion torque model}
How representative are the statistical uncertainties that we get by fitting the the pulse phases with the torque model? The answer to this question is quite relevant if one is interested on investigating phenomena such as the correlation between the NS spin-up and the amount of matter accreting onto its surface. Besides the systematics related to the pulse phase delays already discussed in Sec.~\ref{sec:timing}, the other source of uncertainties that can have a relevant impact on the accreting torque model utilised in this work is related to the X-ray flux measurements. 
\\
Our knowledge on how the X-ray flux changes with time is affected basically by two independent factors: i) uncertainties on the flux estimates and ii) sampling of the outburst. The first factor represents the statistical errors from the spectral modelling of the source emission that we estimated to be on average of the order of 10\%. However, at the end of the outburst, when the source fainted out, this value increases up to 30\%. Another aspect that needs to be taken into account is the degree at which we can reconstruct the evolution of the source properties (e.g., flux variations) during the outburst. Ideally, a continuos monitoring of the source would be required, but for several reasons this is difficult to achieve. In practice, we have a limited number of observations unevenly spaced in time that we can combine, and from which we extrapolate the information of interest. This represents an important limitation when dealing with phenomena happening on timescales shorter than the achievable sampling. For the specific case of the torque modelling, we would like to measure the X-ray flux to quantify the torque exerted to the accreting NS. As explained in Sec.~\ref{sec:timing}, we obviated the lack of continuous sampling of the flux by assuming a linear trend between two consecutive measurements. Although this appears as reasonable approximation, it is also a source of uncertainty that needs to be taken into account when discussing the fitting parameters of the torque model. From Eq.~\ref{eq:phin} we can deduce that the influence of the flux uncertainties on the pulse phase delays is not marginal. In the following we will show, e.g. how the pulse phase varies as a consequence of random fluctuations of the flux around a mean value. To do that, let us consider a generic time interval  $\Delta t_i=t_{i+1}-t_{i}$ in which we define the pulse phase delay as $\Delta \phi_i=\phi_{i+1}-\phi_{i}$. Let us assume that a generic pulse phase can be expressed with respect to a reference time (that we assume to be 0) as $\phi_{i}=\phi_{0}+\nu_0\,t_i+\frac{1}{2}\,\dot{\nu}\,t_i^2$, where $\phi_0$ and $\nu_0$ are the phase and spin frequency values at the reference time, respectively, and $\dot{\nu}\propto F^{\beta}$ is the spin frequency derivative. After some simple algebra we can write the pulse phase delay in the time interval as follows:

\begin{equation}
\Delta \phi_i = \nu_0\,\Delta t_i  + \frac{1}{2}\,\dot{\nu}\,\Delta t_i^2 + \dot{\nu}\,t_i\,\Delta t_i
\end{equation}
\\
and we can estimate the phase fluctuation as:
 
\begin{equation}
\begin{split}
\delta (\Delta \phi_i) &= \left[ \left(\frac{1}{2}\,\delta\dot{\nu}\,\Delta t_i^2\right)^2 + \left(\delta\dot{\nu}\,t_i\,\Delta t_i\right)^2 \right]^{1/2} \approx\\
&\approx \delta\dot{\nu}\,\Delta t_i\,t_i\approx\delta F^{\beta}\Delta t_i\,t_i \approx \left|\frac{F_{i+1}-F_{i}}{2}\right|\Delta t_i\,t_i
\end{split}
\label{eq:rand}
\end{equation}
\\
taking into account that $\delta \nu_0 \simeq 0$ and considering values of $t_i > 2\Delta t_i$. It is worth noting that, although random fluctuations of the flux around a mean value imply a constant term in Eq.~\ref{eq:rand}, the induced pulse fluctuations increase with time, increasing the uncertainties on the accreting torque model used to fit the pulse phase delays. For this reason we decided to investigate this issue by means of Monte Carlo simulations.

\bibliographystyle{aa}
\bibliography{biblio.bib}

\label{lastpage}

\end{document}